# Experimental characterization of the electronic structure of anatase TiO$_2$: Thermopower modulation

Yuki Nagao[1], Akira Yoshikawa[1], Kunihito Koumoto[1], Takeharu Kato[2], Yuichi Ikuhara[2,3], and Hiromichi Ohta[1,4,a]

[1]*Graduate School of Engineering, Nagoya University, Furo-cho, Chikusa, Nagoya 464-8603, Japan*
[2]*Nanostructures Research Laboratory, Japan Fine Ceramics Center, 2–4–1 Mutsuno, Atsuta, Nagoya 456–8587, Japan*
[3]*Institute of Engineering Innovation, The University of Tokyo, 2–11–16 Yayoi, Bunkyo, Tokyo 113–8656, Japan*
[4]*PRESTO, Japan Science and Technology Agency, Sanbancho, Tokyo 102–0075, Japan*

**Thermopower ($S$) for anatase TiO$_2$ epitaxial films ($n_{3D}$: $10^{17}$–$10^{21}$ cm$^{-3}$) and the gate voltage ($V_g$) dependence of $S$ for thin film transistors (TFTs) based on TiO$_2$ films were investigated to clarify the electronic density of states (DOS) around the conduction band bottom. The slope of the $|S|$–log $n_{3D}$ plots was –20 μVK$^{-1}$, which is an order magnitude smaller than that of semiconductors (–198 μVK$^{-1}$), and the $|S|$ values for the TFTs increased with $V_g$ in the low $V_g$ region, suggesting that the extra tail states are hybridized with the original conduction band bottom.**

Anatase TiO$_2$ has received increased attention for several optoelectronic applications, including photocatalysts[1] and dye-synthesized solar cells.[2] Because most of these types of optoelectronic functions in TiO$_2$ depend on the electronic density of states (DOS) around the bandgap between the O 2p valence band and the Ti 3d conduction band,[3] numerous studies have examined the DOS around the bandgap.[4–11] To experimentally clarify the DOS, X-ray and/or ultraviolet photoelectron spectroscopy is used for the valence band, while inverse photoelectron spectroscopy (IPES) is used for the conduction band.[4–8] However, clarifying the DOS around the conduction band bottom is very difficult due to the insufficient energy resolution of IPES (~100 meV[12]). Therefore, an *ab initio* band calculation is mainly used to clarify the DOS around the conduction band bottom for anatase TiO$_2$.[9–11]

In the present study, we aimed to clarify the DOS around the conduction band bottom by examining the thermopower ($S$) of anatase TiO$_2$ epitaxial films. The $S$ value is an effective physical property because it reflects the energy differential of the DOS at the Fermi energy, $[\partial \text{DOS}(E)/\partial E]_{EF}$. Furthermore, the $S$ value depends on the three-dimensional carrier concentration ($n_{3D}$). In particular, a field effect transistor structure is appropriate to measure the $S$ values of semiconductors with different $n_{3D}$ because the gate voltage ($V_g$) can modulate $n_{3D}$ of a FET.[13–15] Although many studies have reported powder,[16] film growth[17,18] and FET fabrication[19,20] of anatase TiO$_2$, few have examined the $S$ value of anatase TiO$_2$.[21–23]

Herein we report the $S$ values for anatase TiO$_2$ epitaxial films ($n_{3D}$: 10$^{17}$–10$^{21}$ cm$^{-3}$) and the $V_g$ dependence of $S$ for a thin film transistor (TFT, on/off ratio >10$^4$, field effect mobility ~0.9 cm$^2$V$^{-1}$s$^{-1}$) based on an undoped TiO$_2$ epitaxial film. We found that the slope of the |$S$|–log $n_{3D}$ plots for anatase TiO$_2$ films is ~ –20 μVK$^{-1}$, which is an order magnitude smaller than that of semiconductors with a parabolic DOS (–198 μVK$^{-1}$= –ln 10·$k_B/e$), and |$S$| values for TFTs increase with $V_g$ in the low $V_g$ region,

suggesting that the extra tail states are hybridized with the original conduction band bottom.

We fabricated several anatase TiO$_2$ epitaxial films by pulsed laser deposition (PLD, KrF excimer laser, 20 ns, ~1 Jcm$^{-2}$pulse$^{-1}$, 10 Hz) on stepped (001) LaAlO$_3$ substrates at 700°C using Nb-doped or undoped rutile TiO$_2$ ceramic as targets. Reflection high energy electron diffraction, high resolution X-ray diffraction, and topographic AFM studies of the resultant films revealed that highly 00l oriented epitaxial TiO$_2$ (anatase) films were obtained with an epitaxial relationship (001)[100]TiO$_2$ ∥ (001)[100] LaAlO$_3$. The TiO$_2$ film thicknesses, which were measured by grazing incidence X-ray reflection, were ~100 nm. Figure 1 shows a cross sectional high-resolution transmission electron microscope (HRTEM) image of a TiO$_2$ film grown on (001) LaAlO$_3$, confirming heteroepitaxial growth of TiO$_2$. An abrupt heterointerface was observed between the TiO$_2$ film and the LaAlO$_3$ substrate.

Then we measured the carrier concentration ($n_{3D}$) and Hall mobility ($\mu_{Hall}$) of the TiO$_2$ films [Fig. 2(a)], which were fabricated using undoped or Nb-doped TiO$_2$ ceramic targets by the dc four probe method with a van der Pauw electrode configuration at room temperature (RT). $\mu_{Hall}$ was independent of $n_{3D}$ (~7 cm$^2$V$^{-1}$s$^{-1}$, which corresponds well to the literature value ~10 cm$^2$V$^{-1}$s$^{-1}$) in the lower $n_{3D}$ region (undoped films, $n_{3D}$<10$^{20}$ cm$^{-3}$). However, an $n_{3D}$ dependence (~$n_{3D}^{-2/3}$) appeared in the higher $n_{3D}$ region (Nb-doped films, $n_{3D}$>10$^{20}$ cm$^{-3}$), indicating that ionized impurity scattering predominantly occurs in the higher $n_{3D}$ region.

Figure 2b shows the |$S$|–log $n_{3D}$ plots for the TiO$_2$ films, which were measured using the conventional steady state method at RT. The negative $S$ values indicated the films have an $n$-type conductivity. The |$S$| value decreased almost proportionally from 150 to 20 μVK$^{-1}$ with log $n_{3D}$. We calculated the |$S$|–log $n_{3D}$ relationship for anatase TiO$_2$ based on the assumption that the DOS around the conduction band bottom is

parabolic with a density of state effective mass of 1.25 $m_0$ where $m_0$ is the free electron mass, for comparison.[24] Although the slope of the calculated $|S|$–log $n_{3D}$ line ($n_{3D}$<$10^{20}$ cm$^{-3}$) corresponded to $-\ln 10 \cdot k_B/e$ (= $-198$ µVK$^{-1}$), that of the observed line was an order magnitude smaller ($-20$ µVK$^{-1}$). In the higher $n_{3D}$ region ($n_{3D}$>$10^{20}$ cm$^{-3}$), the observed $|S|$ values corresponded well to the calculated line. These results clearly indicate that the DOS around the conduction band bottom of the TiO$_2$ epitaxial film is not parabolic.

To further clarify the DOS shape around the conduction band bottom, we subsequently fabricated a top-gate TFT using a TiO$_2$ anatase epitaxial film (undoped, 30 nm thick). First, metallic Ti films (20 nm thick) were deposited for use as the source and drain electrodes. Deposition was performed through a stencil mask by electron beam (EB) evaporation (base pressure ~$10^{-4}$ Pa, no substrate heating). Second, a 200 nm thick Y$_2$O$_3$ film (polycrystalline, dielectric permittivity $\varepsilon_r$=20) was deposited through a stencil mask by PLD (~3 Jcm$^{-2}$pulse$^{-1}$, oxygen pressure ~1 Pa) using a dense polycrystalline Y$_2$O$_3$ ceramic as the target. Finally, a gate electrode, which was a metallic Ti film (20 nm thick), was deposited through a stencil mask by EB evaporation. The resultant TFT was annealed at 200°C in air to reduce the oxygen defects generated during the Y$_2$O$_3$ deposition.

The transistor characteristics of the resultant TiO$_2$ TFT were measured with a semiconductor device analyzer (B1500A, Agilent Technologies) at RT. The channel length ($L$) and channel width ($W$) of the TFT were both 400 µm. Figures 3(a) and (b) show typical transistor characteristics of a TiO$_2$–TFT [(a) drain current ($I_d$)–drain voltage ($V_d$) and (b) $I_d$–$V_g$ curves]. Figure 3(a) clearly shows the current saturation and pinch off behavior, indicating that TFTs obey standard field effect theory. $I_d$ of the TFT increased as $V_g$ increased; hence, the channel was $n$-type and electron carriers were accumulated by positive $V_g$ [Fig. 3(b)]. The on–off current ratio was >$10^4$. The calculated threshold voltage was –17 V from the $I_d^{0.5}$–$V_g$ plot [inset of (b)]. The $\mu_{FE}$ values were obtained from $\mu_{FE}=g_m[(W/L)C_i\cdot V_d]^{-1}$ where $g_m$ is transconductance $\partial I_d/\partial V_g$ and $C_i$ is the capacitance per unit area (89 nFcm$^{-2}$). The $\mu_{FE}$ values of this TFT increased with $V_g$, and reached ~0.9 cm$^2$V$^{-1}$s$^{-1}$ (Fig. 4).

We then measured the field-modulated $S$ of the TiO$_2$ TFT using two Peltier devices to introduce a temperature difference ($\Delta T$ up to 3 K) between the source and drain electrodes. Details of the thermopower measurement are described elsewhere.[14,15] Figure 4 shows the $|S|$–$V_g$ plots for the TiO$_2$ TFT. The negative $S$ values confirmed the channel is an $n$-type. The $|S|$ value gradually increased from 330 to 390 µVK$^{-1}$ as $V_g$ increased in the lower $V_g$ region ($V_g$<12 V), and the sheet carrier concentration increased. On the other hand, in the higher $V_g$ region ($V_g$>12 V), the $|S|$ value gradually decreased with $V_g$. As discussed above, the $|S|$ values should decrease with $V_g$ because the carrier concentration increases with $V_g$ for a parabolic DOS shape around the conduction band bottom. Because a small transition of $I_d$ was also observed at $V_g$~12 V in the $I_d$–$V_g$ curve [Fig. 3(b)], the electronic states of the channel returned to the original DOS parabolic shape upon applying $V_g$>12 V.

These results indicate the DOS around the conduction band bottom for a TiO$_2$ epitaxial film is composed of not only the original parabolic DOS (high $\mu$), but also small non-parabolic shaped DOS such as a tail state (low $\mu$) hybridized near the conduction band bottom.

In summary, we have shown that the thermopower ($S$) measurements for anatase TiO$_2$ epitaxial films (undoped and Nb-doped, carrier concentration, $n_{3D}$: 10$^{17}$–10$^{21}$ cm$^{-3}$) are effective to experimentally clarify the electronic density of states (DOS) near the conduction band bottom. The slope of the $|S|$–log $n_{3D}$ plots for anatase TiO$_2$ films is –20 µVK$^{-1}$, which is an order magnitude smaller than that of semiconductors with parabolic DOS (–198 µVK$^{-1}$= ln 10·$k_B$/$e$), and the $|S|$ values for TFTs increase with $V_g$ in the low $V_g$ region. The DOS around the conduction band bottom for an TiO$_2$ epitaxial film is composed of both the parabolic shaped original DOS (high $\mu$) and a small non-parabolic shaped DOS such as a tail state (low $\mu$) hybridized near the conduction band bottom.

Some of this work was financially supported by MEXT (22360271, 22015009).

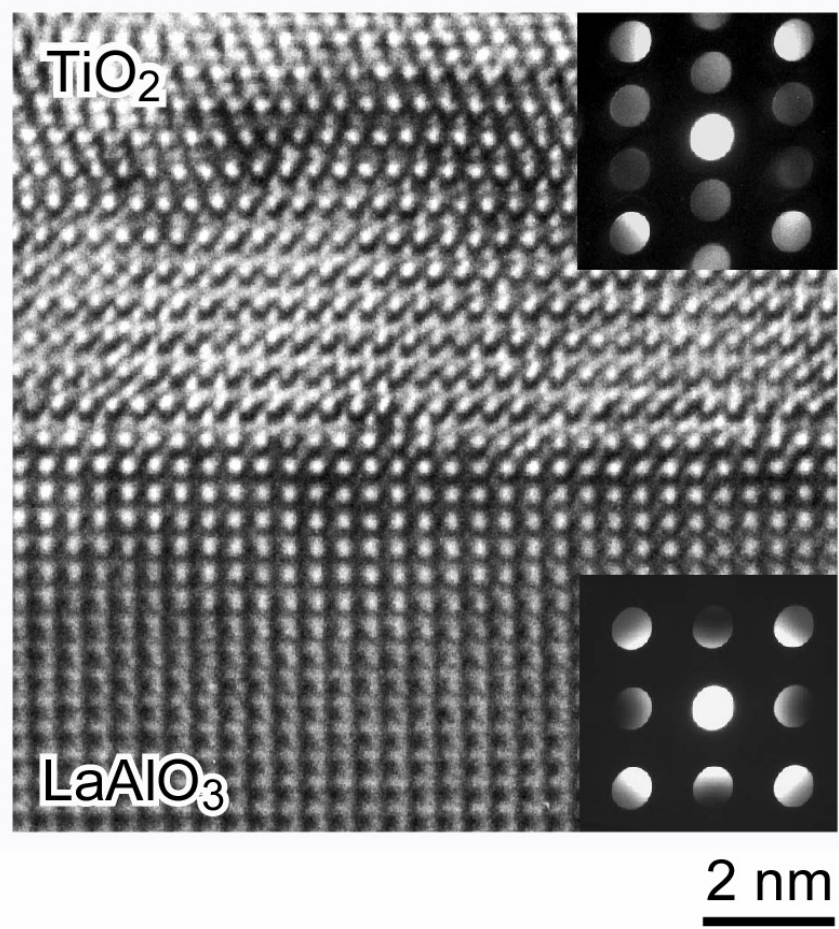

**Fig. 1** Cross sectional HRTEM image of the TiO$_2$ epitaxial film grown on a (001) LaAlO$_3$ substrate with an epitaxial relationship (001)[100]TiO$_2$ || (001)[100] LaAlO$_3$. Nanobeam electron diffraction patterns for the TiO$_2$ and LaAlO$_3$ are also shown.

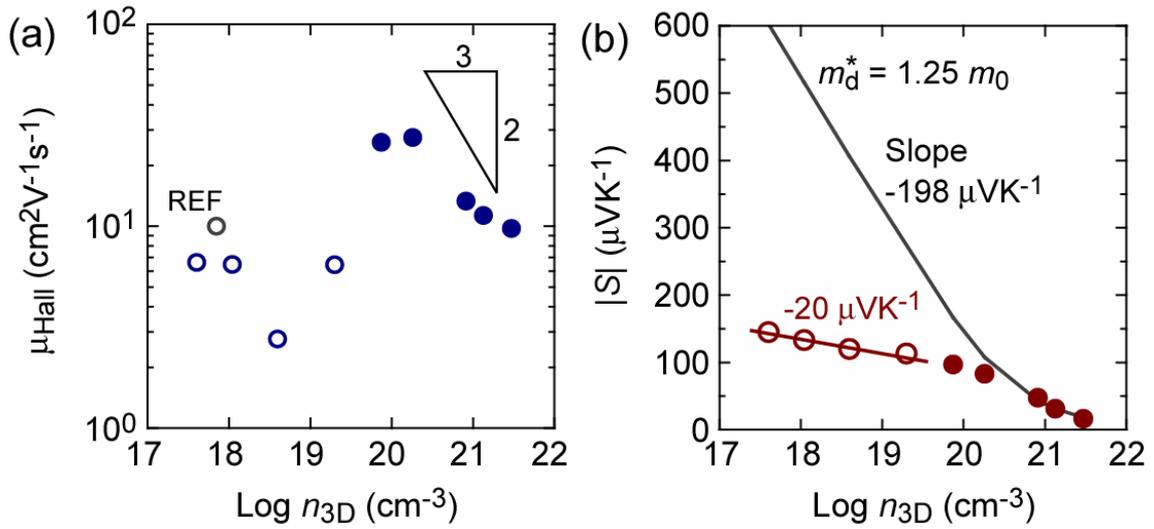

**Fig. 2** Carrier concentration ($n_{3D}$) dependence of (a) Hall mobility ($\mu_{Hall}$) and (b) thermopower ($|S|$) for $TiO_2$ epitaxial films (undoped and Nb-doped, ~100 nm thick) grown on (001) $LaAlO_3$. For comparison, $\mu_{Hall}$ of a bulk single crystal (REF 21) is plotted in (a) and the calculated $|S|$–log $n_{3D}$ relationship for anatase $TiO_2$ is shown in (b).

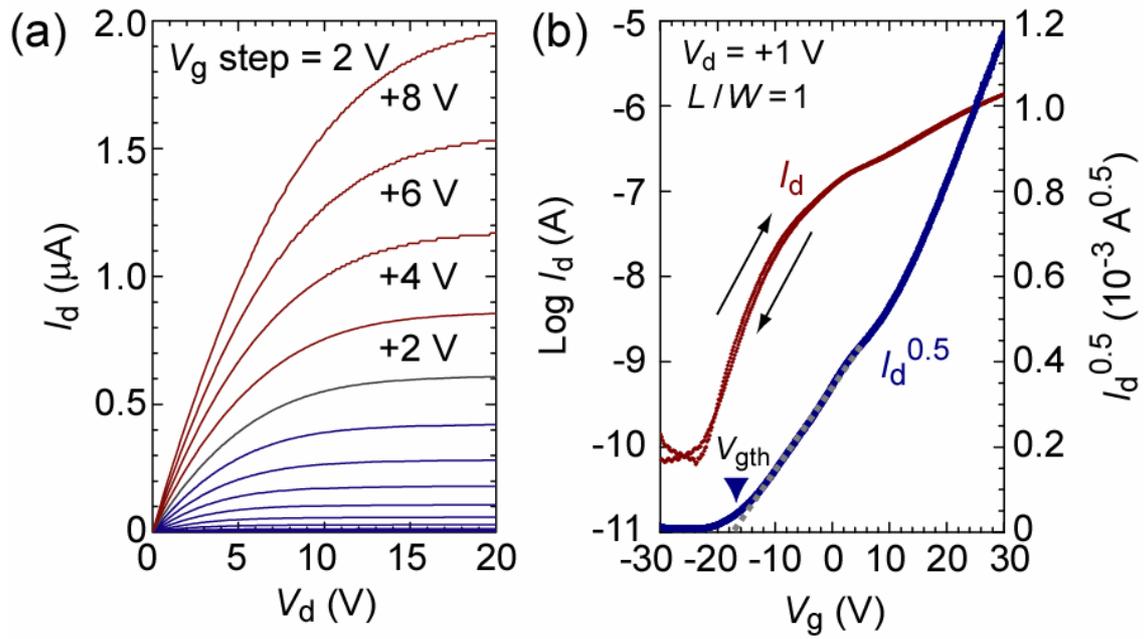

**Fig. 3** Typical transistor characteristics [(a) $I_d$–$V_d$ and (b) $I_d$–$V_g$ curves] of a TiO$_2$ TFT at RT. Current saturation and pinch off behavior are clear in (a). $I_d^{0.5}$–$V_g$ plots are shown in (b). On–off current ratio is >10$^4$ and $V_{gth}$ is −17 V.

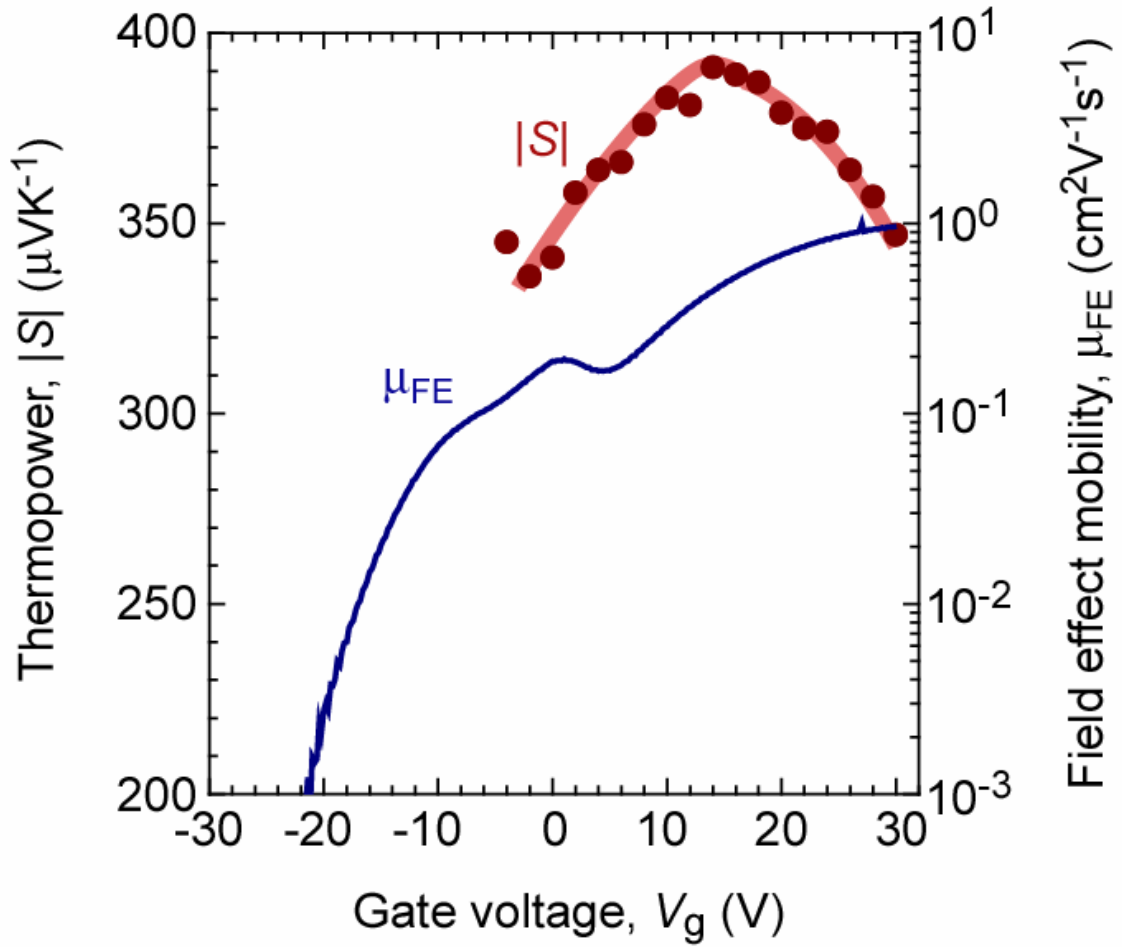

**Fig. 4** Field modulation of thermopower, |S|–$V_g$ plots of TiO$_2$ TFTs at RT. $\mu_{FE}$–$V_g$ plots are also shown. As $V_g$ increases in the lower $V_g$ region ($V_g$ < 12 V), the |S| value gradually increases from 330 to 390 µVK$^{-1}$, whereas the |S| value gradually decreases with $V_g$ in the higher $V_g$ region ($V_g$ > +12 V). As $V_g$ increases, $\mu_{FE}$ gradually increases.